\pdfoutput=1
\documentclass[a4paper]{article}

\usepackage[boxruled]{algorithm2e}
\SetKw{Continue}{continue}
\SetKw{Break}{break}
\SetKw{True}{true}

\usepackage{amsmath}
\usepackage{amsfonts}

\usepackage{amssymb}
\usepackage{tikz}
\usepackage{mathrsfs}
\usepackage[colorinlistoftodos]{todonotes}

\usepackage{xcolor}
\usepackage{hyperref}
\hypersetup{
    colorlinks,
    linkcolor={green!50!black},
    citecolor={blue!50!black},
    urlcolor={blue!80!black}
}

\usepackage[inline]{enumitem}

\usepackage{amsthm}
\usepackage{cases}

\usepackage[capitalize]{cleveref}

\newtheorem{theorem}{Theorem}
\newtheorem{lemma}[theorem]{Lemma}
\newtheorem{corollary}[theorem]{Corollary}
\newtheorem{definition}[theorem]{Definition}

\newtheorem{remark}[theorem]{Remark}
\newtheorem{fact}[theorem]{Fact}
\newtheoremstyle{restate}{}{}{\itshape}{}{\bfseries}{~(restated).}{.5em}{\thmnote{#3}}
\theoremstyle{restate}

\DeclareMathOperator*{\argmax}{arg\,max}

\newcommand{\cD}{\mathcal{D}}
\newcommand{\cE}{\mathcal{E}}
\newcommand{\cF}{\mathcal{F}}
\newcommand{\cB}{\mathcal{B}}

\DeclareMathOperator{\Unif}{Unif}
\DeclareMathOperator*{\E}{\mathbb{E}}
\newcommand{\eps}{\epsilon}
\newcommand{\R}{\mathbb{R}}
\newcommand{\Norm}[1]{\left\| #1 \right\|}

\newcommand{\inner}[2]{\left\langle #1, #2 \right\rangle}
\newcommand{\match}{\leftrightarrow}

\usepackage{stackengine}

\newcommand{\ee}{\stackon[2pt]{\^{e}}{\~{}}}

\DeclareMathOperator{\polylog}{polylog}
\DeclareMathOperator{\nnz}{nnz}

\usepackage[margin=1in]{geometry}

\title{Lower Bounds for Sparse Oblivious Subspace Embeddings} %

\author{Yi Li\thanks{Supported in part by Singapore Ministry of Education (AcRF) Tier 2 grant MOE2018-T2-1-013.}\qquad\qquad Mingmou Liu\thanks{Supported by Singapore Ministry of Education (AcRF) Tier 2 grant MOE2018-T2-1-013.}\\
School of Physical and Mathematical Sciences\\
Nanyang Technological University\\
\texttt{\{yili,mingmou.liu\}@ntu.edu.sg}
}

\date{}

\begin{document}

\maketitle

\begin{abstract}
An oblivious subspace embedding (OSE), characterized by parameters $m,n,d,\eps,\delta$, is a random matrix $\Pi\in \R^{m\times n}$ such that for any $d$-dimensional subspace $T\subseteq \R^n$, $\Pr_\Pi[\forall x\in T, (1-\eps)\|x\|_2 \leq \|\Pi x\|_2\leq (1+\eps)\|x\|_2] \geq 1-\delta$. For $\eps$ and $\delta$ at most a small constant, we show that any OSE with one nonzero entry in each column must satisfy that $m = \Omega(d^2/(\eps^2\delta))$, establishing the optimality of the classical Count-Sketch matrix. When an OSE has $1/(9\eps)$ nonzero entries in each column, we show it must hold that $m = \Omega(\eps^{O(\delta)} d^2)$, improving on the previous $\Omega(\eps^2 d^2)$ lower bound due to Nelson and Nguy\ee{}n (ICALP 2014).
\end{abstract}

\section{Introduction}

Subspace embedding is a fundamental technique in dimensionality reduction which has been successfully applied to a variety of tasks on large data sets, including clustering~\cite{BZMD,Cohen15}, correlation analysis~\cite{ABTZ} and typical linear algebraic problems such as regression and low-rank approximation~\cite{CW17}. In general, the subspace embedding technique ``compresses'' a large-scale problem, which is computationally prohibitive, into a smaller one so that the result of the original problem can be obtained from the compressed version and the computation on the compressed version is affordable. A survey of subspace embedding techniques and their applications can be found in~\cite{W14}.

More specifically, we define the notion of \emph{oblivious subspace embedding} as follows.
\begin{definition}
An $(n,d,\eps,\delta)$-oblivious-subspace-embedding is a random matrix $\Pi\in \R^{m\times n}$ such that for any fixed $d$-dimensional subspace $T\subseteq \R^n$, it holds that
\begin{equation}\label{eqn:OSE}
\Pr\left\{\forall x\in T, (1-\eps)\|x\|_2 \leq \|\Pi x\|_2 \leq (1+\eps)\|x\|_2\right\} \geq 1 - \delta.
\end{equation}
\end{definition}
When $n$ and $d$ are clear from the context, we abbreviate an $(n,d,\eps,\delta)$-oblivious-subspace-embedding to an $(\eps,\delta)$-oblivious-subspace-embedding. Very often, the subspace $T$ is the column space of a matrix $A\in \R^{n\times d}$ and the subspace embedding property~\eqref{eqn:OSE} thus becomes 
\begin{equation}\label{eqn:SE_matrix}
\Pr \left\{\forall x\in \R^d, (1-\eps)\|Ax\|_2 \leq \|\Pi A x\|_2 \leq (1+\eps)\|Ax\|_2\right\} \geq 1 - \delta.
\end{equation}

A typical construction of $\Pi$ is a random Gaussian matrix of i.i.d.\ $N(0,1/m)$ entries with $m = \Theta((d + \log(1/\delta))/\eps^2)$. 
This target dimension $m$ is known to be optimal up to a constant~\cite{NN14}. However, it is a dense matrix and computing $\Pi A$ could be 
computationally expensive. As a result, having a sparse matrix $\Pi$, which is commonly characterized by the maximum number of nonzero entries in each column, denoted by $s$, is highly desirable. OSNAP~\cite{NN13:OSNAP,Cohen16} and Count-Sketch~\cite{CW17} are two classical sparse constructions. In the OSNAP construction, $m=\Theta(d \log(d/\delta) /\eps^2)$ and $\Pi$ has $s=\Theta(\log(d/\delta)/\eps)$ nonzeros per 
column, whose positions are chosen uniformly in each column and whose values are Rademacher variables; alternatively, $m=\Theta(d^{1+\gamma}\log(d/\delta)/\eps^2)$ and $s=\Theta(1/(\gamma\eps))$ for any constant $\gamma>0$. The column sparsity allows for 
computing $\Pi A$ in $O(\nnz(A)\cdot s)$ time, where $\nnz(A)$ denotes the number of nonzero entries of $A$. The Count-Sketch 
construction is a special and extreme case of the OSNAP construction in which the column sparsity parameter $s=1$. It has, however, $m = \Theta(d^2/(\delta\eps^2))$ rows, which is quadratic in $d/\eps$. Despite this, the matrix $\Pi$ is extremely sparse and computing $\Pi A$ 
takes only $O(\nnz(A))$ time, making Count-Sketch indispensable in several fastest algorithms. 
A natural question is, therefore, 

\begin{center}\it
Is it possible to further reduce the target dimension $m$ in the sparse constructions of $\Pi$?
\end{center}

To answer the question, one can consider the lower bound for the target dimension $m$, that is, to show that an oblivious subspace embedding cannot exist if $m$ is less than a certain threshold. So far only a few results have been known in this direction. 
Nelson and Nguy\ee{}n proved that for $s=1$ and constant $\eps,\delta$, any $(\eps,\delta)$-oblivious-subspace-embedding must have $m=\Omega(d^2)$~\cite{NN13:LB}, showing the tightness of the quadratic dependence on $d$ for Count-Sketch. When combined with the general (unconstrained column sparsity) lower bound of $m = \Omega(d/\eps^2)$, the best previously known lower bound for the case of $s=1$ is $\Omega(d^2 + d/\eps^2)$, assuming that $\delta$ is a constant. This does not preclude a smaller dimension for Count-Sketch or, in general, extreme sparse constructions with $s=1$.
For larger $s$, the same authors showed that for $\delta$ a constant and $n = \Omega(d^2)$, any $(\eps,\delta)$-oblivious-subspace-embedding must have $m = \Omega(\eps^2 d^2)$ when $s \leq \alpha/\eps$ for some constant $\alpha>0$~\cite{NN14}. Recall the general $\Omega(d/\eps^2)$ bound without the sparsity constraint, the $\Omega(\eps^2 d^2)$ bound is stronger only when $d=\Omega(1/\eps^4)$. We also note that the sparsity constraint $s\leq \alpha/\eps$ is optimal up to a constant factor for the quadratic dependence on $d$ due to the aforementioned OSNAP construction, where $s=\Theta(1/(\gamma\eps))$ and $m=\Theta(d^{1+\gamma}(\log d)/\eps^2)$. 

\paragraph{Our contributions.} We obtain the following improvements on the lower bounds on $m$ for any $(\eps,\delta)$-oblivious-subspace-embedding.
\begin{enumerate}[label=(\roman*)]
\item When $s=1$, we show a lower bound of $m = \Omega(d^2/(\eps^2\delta))$ for $n = \Omega(d^2/(\eps^2\delta))$, which implies that the target dimension of Count-Sketch is tight up to a constant factor.
\item When $s\leq 1/(9\eps)$, we show a lower bound of $m = \Omega(\eps^{\Theta(\delta)} d^2)$ for $d=\Omega(1/\eps^2)$, $n=\Omega(d^2/\eps^2)$ and constant $\delta$. This is almost a quadratic improvement on dependence on $\eps$ compared with the previous lower bound of $\Omega(\eps^2 d^2)$. 
 Our result also reduces the effective constraint $d=\Omega(1/\eps^4)$ to $d=\Omega(1/\eps^{2+O(\delta)})$, again almost a quadratic improvement. Furthermore, we obtain a trade-off between $m$ and $s$, namely $m = \Omega((\log s)^{-\Theta(1)}s^{-\Theta(\delta)}d^2)$, when $s\leq 1/(9\eps)$.
\end{enumerate}

\subsection{Our Techniques} 

We give a brief overview of the techniques for our new lower bounds, assuming that $\delta$ is a small constant. 
First of all, to show a lower bound of $m \geq M(d,\eps,\delta)$ for an $(\eps,\delta)$-oblivious-subspace-embedding, it suffices, by Yao's minimax principle, to construct a distribution $\cD$ over $n\times d$ matrices (called the hard instance) and show that if a deterministic matrix $\Pi\in\R^{m\times n}$ is an $(\eps,\delta)$-subspace-embedding for $A\sim \cD$, i.e., if \eqref{eqn:SE_matrix} holds for probability taken over $A\sim \cD$, then $\Pi$ must have at least $M(d,\eps,\delta)$ rows. Without loss of generality, we may assume that $\cD$ is supported on isometries. Here, an isometry $U\in\R^{n\times d}$ has orthonormal columns. 

\subsubsection{Tight lower bound for $s=1$}
We first review an easy proof of the $m = \Omega(d^2)$ lower bound for $s=1$. Intuitively, if $\Pi$ is a subspace embedding for a deterministic isometry $U$, there cannot be a row $l$ and two columns $i,j$ of $\Pi U$ such that $|(\Pi U)_{l,i}|$ and $|(\Pi U)_{l,j}|$ are both large, because the ``collision'' would make the norm $\Norm{\Pi U u}_2$ not concentrate around $1$ for some unit vector $u\in\R^d$.
For example, consider the simple case of $s=1$, $V=(I_d\quad 0)^T$ and $U = S\cdot V$, where $S\in\R^{n\times n}$ is a diagonal matrix of i.i.d.\ Rademacher variables on its diagonal line, and suppose that $|(\Pi U)_{l,i}|, |(\Pi U)_{l,j}|\in [1-\epsilon,1+\epsilon]$. 
Then for $u\triangleq (e_i+e_j)/\sqrt{2}$ we have $\Norm{\Pi U u}_2^2\geq 2(1-\eps)^2$ with probability $1/2$ and $\Norm{\Pi U u}_2^2 \le 2\eps^2$ with $1/2$. Observe that $2(1-\eps)^2$ and $2\eps^2$ are $2-4\eps$ apart from each other, which contradicts the fact that $\Norm{\Pi U u}_2^2$ falls in $[(1-\epsilon)^2,(1+\epsilon)^2]$, an interval of length $4\epsilon < 2-4\eps$.\footnote{In the case of $s=1$, Nelson and Nguy\ee{}n proved $m=\Omega(d^2)$ by observing that the rank of $\Pi U$ would be less than $d$ if a ``collision'' exists~\cite{NN13:LB}. Their argument seems difficult to apply to more complicated hard instances, while our type of contradiction is more amenable and plays a central role in proving higher lower bounds in our work.}
Based on this observation, in the case of $s=1$, one can choose the hard instance $U$ to be $U=SV$, where $S$ is as before but $V$ is the induced distribution by randomly permuting the rows of $(I_d\quad 0)^T$. Since $\Pi U$ behaves like hashing $d$ coordinates into $m$ buckets and the subspace embedding property implies that there should be no collision with a large probability, the lower bound of $m=\Omega(d^2)$ follows easily from the birthday paradox.

To obtain a better lower bound for $s=1$, the hard instance $U$ has to capture more than $d$ dimensions in its rows. Hence, we replicate the identity matrix $1/(8\epsilon)$ times and normalize it, that is, let $V$ be a row-permuted version of $(I_d\quad I_d\quad \cdots\quad I_d\quad 0)^T$ and $U = \sqrt{8\eps}SV$, so that every column of $U$ has exactly $1/(8\epsilon)$ nonzero entries, each being $\pm\sqrt{8\epsilon}$.
Similarly to before, we seek to prove that no collisions should exist. Here, a collision means that $\Pi$ has large values at some row $l\in[m]$ and some columns $i',j'\in[n]$ and, simultaneously, $U_{i',i},U_{i',j}\neq 0$.\footnote{Throughout this paper, we let $[n]\triangleq\{1,\dots,n\}$.} If there ever exists a ``collision'', we would like to argue that we can find a unit vector $u$ and a real number $o$ such that $\Norm{\Pi U u}_2^2 > o + 2\epsilon$ with probability at least $1/4$ and $\Norm{\Pi U u}_2^2 < o - 2\epsilon$ with probability at least $1/4$ (over the randomness of $S$), and thus reach a contradiction.
We still set $u\triangleq (e_i+e_j)/\sqrt{2}$.
However, an important difference from before, thus the main difficulty, is that $\Pi U u$ now may have nonzero coordinates other than the $l$-th coordinate since $U$ has more than one nonzero entry in each column, and so $\Norm{\Pi U u}_2^2$ cannot be easily controlled by observing the $l$-th coordinate with an easy bound or a concentration bound for the other coordinates. Even the $l$-th coordinate $(\Pi U u)_l$ itself may contain a hard-to-control contribution from columns other than columns $i',j'$ in $\Pi$ since there may exist $k'\neq i',j'$ such that $\Pi_{l,k'}U_{k',i}\neq 0$.
Our key observation is the following. 
Conditioning on that $\Pi_{l,i'},\Pi_{l,j'},U_{i',i},U_{j',j}\ne 0$, for any fixing of the random variables other than $S_{i',i'},S_{j',j'}$,
we can find a real number $o$ such that $\Norm{\Pi U u}_2^2 > o + 2\epsilon$ with probability $1/4$ and $\Norm{\Pi U u}_2^2 < o - 2\epsilon$ with probability $1/4$. 
This suffices to derive a contradiction. 
Hence, all $d/(8\eps)$ nonzero entries of $U$ must be isolated under the hashing matrix $\Pi$ and the lower bound $m = \Omega(d^2/\eps^2)$ follows again from the birthday paradox.

\subsubsection{New lower bound for $s\leq 1/(9\eps)$}

\paragraph{Generalizing hard distribution.}
We generalize the hard instance $U$ above with parameter $\beta$. Let $\cD_\beta$ be the distribution of random matrices $U$ containing $1/\beta$ copies of $I_d$, normalized by $\sqrt{\beta}$ and multiplied by independent Rademacher variables, that is, $V$ is a row-permuted version of $(I_d\quad I_d\quad \cdots\quad I_d\quad 0)^T$ and $U \triangleq \sqrt{\beta}SV$. A formal definition can be found in Definition~\ref{def:D_beta}.\footnote{The distribution of $U$ in Definition~\ref{def:D_beta} is not strictly supported on isometries but is an isometry with an overwhelming probability, which suffices for our purpose.}
We observe that OSNAP with column sparsity $s\le 1/(8\epsilon)$ requires $m=\Omega(d^2/\epsilon^2)$ rows for $U\sim\cD_{8\epsilon s}$. This suggests the following conjecture: if $\Pi$ has many entries of large absolute value, then $\cD_1$ is a hard distribution. %
Therefore, we start by proving a lower bound for the hard instance $U\sim\cD_1$ under an ``abundance assumption'' that $\Pi$ has $\Omega(n/\eps)$ entries of absolute value $\Omega(\sqrt{\epsilon})$, and remove the assumption later at the cost of losing an $\epsilon^{O(\delta)}$ factor in the lower bound.

\paragraph{Generalizing the collision.}
To handle the case $s>1$, we generalize the notion of collision of large entries.
Instead of looking for two large entries in the same row, we look for two columns with a large inner product in absolute value.
Assuming the large inner product, we can show, using an argument analogous to the case $s=1$, a contradiction to the assumption that $\Norm{\Pi U u}_2^2$ is contained in an interval of length $4\eps$ for some unit vector $u$ (See Lemma~\ref{lemma: inner product implies anticoncentration} for a precise statement).
Therefore, it suffices to show that, if $m$ is small, with constant probability, there exist two columns of $\Pi$ chosen by $U$ whose inner product is at least $5\eps$ in absolute value.
Observe that the collision of two entries of absolute value at least $\sqrt{8\epsilon}$ contributes at least $8\epsilon$ to the inner product.
Thus, it suffices to prove that the inner product among the remaining entries does not cancel the $8\eps$ by more than $3\epsilon$. To this end,
we adopt an idea from Nelson and Nguy\ee{}n \cite{NN14} and show that the inner product between two columns, after removing the two large colliding entries, is close to zero with good probability.
Now, if we assume that $m=O(d^2)$, $\Pi$ has sufficiently many entries of absolute value $\Omega(\sqrt{\epsilon})$ and $U\sim\cD_1$, there will be many collisions of large entries with constant probability. 
For each such collision, as previously argued, with good probability, the inner product of the corresponding columns has absolute value at least $5\eps$. 
Therefore, with constant probability, we can find a pair of columns of $\Pi$ chosen by $U$ such that their inner product has absolute value at least $5\eps$, assuming that $m$ is small and $\Pi$ has many entries of large absolute values.

\paragraph{Finding colliding column pairs.}
The central work to prove the lower bound for $s=O(1/\epsilon)$, under the ``abundance assumption'', lies in finding and analyzing the distributions of the random column pairs $(i,j)$ of $\Pi$ given the existence of a row $l$ such that $|\Pi_{l,i}|,|\Pi_{l,j}|$ are both large.
In fact, using Nelson and Nguy\ee{}n's idea to obtain a stronger lower bound is highly non-trivial. (One can refer to \Cref{lem: small inner product} for the formal statement of a proposition akin to their idea.) This idea works only when columns $i,j$ are drawn from the same set uniformly without replacement.
However, this may indeed not be satisfied by a random colliding column pair. %
For instance, suppose that $\Pi_{l_1,i}=\Pi_{l_1,j}=\Pi_{l_2,j}=\Pi_{l_2,k}=\sqrt{8\epsilon}$ and the remaining entries are zero, then $(i,j),(j,k)$ are colliding pairs but $(i,k)$ is not.

A simple workaround is to limit the collisions between a column $i$ and other columns to some fixed row $l(i)$.
Let $l(i)$ be such that $\Pi_{l(i),i}$ has the largest absolute value among the $m$ entries in column $i$ and consider only the collisions between column $i$ and columns $j$ where $l(j)=l(i)$.
Under this restriction, the columns are partitioned into disjoint groups based on $l(i)$'s, and the distribution of colliding column pair can be viewed as sampling a group followed by sampling  uniformly at random two columns $i,j$ in the group.
A critical drawback of this solution is that we can merely find $\approx\sqrt{d^2/m}$ colliding column pairs, and each pair yields large inner product with probability $\approx\eps$, so Nelson and Nguy\ee{}n obtained a lower bound of $m=\Omega(\epsilon^2d^2)$ only.

To obtain a higher lower bound, we shall balance 
\begin{enumerate*}[label=(\roman*)]
    \item 
        the number of colliding column pairs we examine and
    \item
        the probability of each pair having a large inner product in absolute value,
\end{enumerate*}
so that we can maximize the probability that we find a pair of columns whose inner product is large in absolute value.
To maximize (i), we adopt the following greedy strategy. Recall that the matrix $U$ in the definition of $\cD_1$ has $d$ nonzero rows at random positions and let $C_1,\dots,C_d\in [n]$ be the indices of these rows. 
Initialize $S\triangleq [d]$. We iterate over $i\in[d]$. If $i\in S$, we find $j\in S\setminus\{i\}$ such that the columns $C_i$ and $C_j$ collide, output a pair $(C_i,C_j)$ and remove $i,j$ from $S$.
Indeed, this greedy strategy would result in a large number of colliding pairs, but the problem is how to show (ii) is large.
A challenge is that the distribution of the column pairs found by the greedy strategy is rather involved.
To see this, let $\phi_c$ denote the conditional probability of a random $C_j$ collides with $C_i$ given that $C_i=c$, then the marginal distribution of $C_i$ of a random colliding pair $(C_i,C_j)$ has probability mass function $p(c)=\phi_c/\sum_c\phi_c$. The greedy strategy, on the other hand, induces a marginal probability $q(c)=\psi_c/\sum_c\psi_c$ for $C_1$, conditioned on that $C_1$ collides with $C_j$ for some $j>1$, where $\psi_c\triangleq 1-(1-\phi_c)^{d-1}$ is the probability that there exists $j>1$ such that $C_j$ collides with $C_1=c$. Thus, the colliding pair found by the greedy strategy has a different distribution from the true distribution $\cD'$ of a colliding pair $(C_1, C_j)$.
To overcome this challenge, we observe that 
\begin{enumerate*}[label=(\roman*)]
    \item if $\phi_c=\Omega(1/d)$ for some $c$, there will exist a row $l$ such that $|\Pi_{l,i}|$ is large for $\Omega(1/(s d))$-fraction of columns $i\in[n]$ and we can then apply Nelson and Nguy\ee{}n's idea on row $l$;
    \item if $\phi_c=O(1/d)$ for all $c$, then $\psi_c=\Theta(d\cdot\phi_c)$ and $q(c)=\Theta(p(c))$ for all $c$, hence the distribution of a colliding column pair found by the greedy strategy is close to the true distribution $\cD'$.
\end{enumerate*}
Assuming that $\phi_c=O(1/d)$ for all $c$, it is sufficient to analyze $\cD'$, which is another technical innovation of this work.
Below we assume that it always holds $\phi_c=O(1/d)$ for all $c$ and analyze the pairs found in this case. 

\paragraph{Analyzing random colliding column pairs.}
Recall that Nelson and Nguy\ee{}n's idea works only when columns $i,j$ are drawn from the same set uniformly without replacement, and $\cD'$ may not be such a distribution.
Our idea is to examine a distribution $\cD''$ which is close to $\cD'$ and is a linear combination of distributions that sample column pairs without replacement. 
Specifically, we assign to each colliding column pair $(i,j)$ a weight which is equal to the number of rows $l$ such that $\Pi_{l,i},\Pi_{l,j}$ are both large in absolute value, and let $\cD''(i,j)$ be proportional to the weight of $(i,j)$.
Under this construction, $\cD''$ can be viewed as sampling a row $l$ from some distribution, followed by sampling two columns $i,j$ uniformly without replacement from the set of large entries in row $l$. Note that for any $(i,j)$, it holds that $\cD'(i,j)=\Omega(\epsilon\cdot \cD''(i,j))$, since the column sparsity $s=O(1/\epsilon)$. 
Invoking Nelson and Nguy\ee{}n's idea, we can show that a colliding column pair $(i,j)\sim\cD''$ has a large inner product with probability $\Omega(\epsilon)$. Consequently, a column pair $(i,j)\sim\cD'$ has a large inner product with probability $\Omega(\epsilon^2)$. 
Since an average colliding column pair can collide in $\Omega(1/\epsilon)$ rows, we have to set $m=O(\epsilon d^2)$ to ensure $\Omega(1/\epsilon^2)$ colliding column pairs, and thus obtain a lower bound of $m=\Omega(\epsilon d^2)$ only.

Our final solution is to further consider a quantity $\Delta$ of $\Pi$, which is the expected number of rows $l$ such that $|\Pi_{l,i}|,|\Pi_{l,j}|$ are simultaneously large for $(i,j)\sim\cD'$.
We take $\Delta$ along with $\cD''$ as a pivot for our proof.
We show that we can find $\Omega(d^2/(\epsilon^2 m\Delta))$ colliding column pairs with high probability and, informally, that a random pair of colliding columns $(i,j)\sim\cD'$ has a large inner product with probability $\Omega(\epsilon^2\Delta)$.
Therefore, if $m\le d^2$, we find a column pair with a large inner product with constant probability, leading to the lower bound $m=\Omega(d^2)$.

\paragraph{Removing the assumption.}
The key argument for removing the ``abundance assumption'' is the following claim. If $m$ is small, then for every $\ell\in \{0,1,\dots,L\}$ ($L\triangleq \log(1/\epsilon)-3$), the matrix $\Pi$ cannot have too many entries of absolute value at least $\sqrt{2^{\ell}}$.
This claim would imply that the $\ell_2$-norm of an average column of $\Pi$ is less than $1-\epsilon$ and thus $\Pi$ cannot be a subspace embedding for $U\sim \cD_1$.

To prove the claim, we propose a new hard distribution: with probability $1/2$, it is $\cD_1$; with probability $1/2$, it is $\cD_{2^{-\ell'}}$ for $\ell'$ uniformly chosen from $\{0,1,\dots,L\}$.
By the law of total probability, $\Pi$ must be a subspace embedding for $U\sim\cD_1$ and also a subspace embedding for $U\sim\cD_{2^{-\ell'}}$ for $(1-O(\delta))$-fraction of $\ell'\in \{0,\dots,L\}$.
Hence, for each $\ell\in \{0,\dots,L\}$, there exists an $\ell'$ among the $(1-O(\delta))$-fraction such that $\Pi$ is a subspace embedding for $U\sim \cD_{2^{-\ell'}}$ and $\ell+\ell'\approx L$ up to an additive error of $O(\delta L) $, matching our conjecture that $U\sim\cD_{2^{-L+\ell}}$ is hard for the subspace embedding $\Pi$ with column sparsity $s=2^{\ell}$. %
We can then follow the proof of the lower bound with the ``abundance assumption'' to show that $\Pi$ cannot have $\approx\eps^{O(\delta)}2^{\ell}n$ entries of absolute value at least $\sqrt{2^{\ell}}$, where the additional $\eps^{O(\delta)}$ factor is incurred by the aforementioned additive error $O(\delta L)$. 
Thus, the average squared $\ell_2$-norm of column vectors of $\Pi$ is $\eps^{O(\delta)}\log(1/\eps) < (1-\eps)^2$ for small enough $\eps$, contradicting the assumption that $\Pi$ is a subspace embedding for $U\sim\cD_1$.

\section{Preliminaries}

For $x,y\in\R$ and $\theta>0$, we write $x= y\pm\theta$ if $x\in[y-\theta,y+\theta]$. For a matrix $A$, we denote its $i$-th column vector by $A_{\ast, i}$. For a finite set $S$, we denote by $\Unif(S)$ the uniform distribution on $S$. For a random variable $X$ and a probability distribution $\cD$, we write $X\sim \cD$ to denote that $X$ follows $\cD$.

When $d$ and $\eps$ are clear from the context, we abbreviate the event in the probability of \eqref{eqn:SE_matrix} to ``$\Pi$ is a subspace embedding for $A$''.

The following distribution $\cD_\beta$ on $n\times d$ matrices, parameterized by $\beta$, is fundamental in the hard instances for the lower bounds.

\begin{definition}[Distribution $\cD_\beta$]\label{def:D_beta}
The distribution $\cD_\beta$ ($0 < \beta\leq 1$) is defined on matrices $U\in\mathbb{R}^{n\times d}$ as follows. 
The matrix $U$ is decomposed as $U=VW$, where $V\in\mathbb{R}^{n\times d/\beta}$ and $W\in\mathbb{R}^{d/\beta\times d}$.
The matrix $V$ has i.i.d.\ columns, each $V_{\ast,i}$ ($i=1,\dots,d/\beta$) is uniformly distributed among the $n$ canonical basis vectors in $\mathbb{R}^n$. %
The matrix $W$ is distributed as follows:
For each $i=1,\dots,d$, set $W_{j,i} \triangleq \sigma_{j}\sqrt{\beta}$ for $j=(i-1)/\beta+1,\dots,i/\beta$, where $\sigma_{j}\in\{-1,1\}$ are independent Rademacher variables; set the remaining entries of $W$ to zero.
\end{definition}
Let $\cB$ denote the event that $V$ has two identical columns.
Since $U=VW$ may not be an isometry when $\cB$ happens, we shall only consider the subspace embedding property of $\Pi$ for $U$ conditioned on $\overline{\cB}$.
Note that when $n\ge K d^2/(\beta^2\delta)$, it holds that $\Pr[\cB]\leq \delta/(2K)$, which can be made an arbitrarily small fraction of $\delta$ by setting $K$ large enough. We shall thus assume $K$ is large enough and ignore the event $\cB$ hereafter and pretend that $V$ has independent columns and the full column rank.

Our strategy is to show that there are two columns $i,j$ of $\Pi V$ such that their inner product $\inner{(\Pi V)_{\ast, i}}{(\Pi V)_{\ast, j}}$ deviates from zero and that the large deviation implies that $\Pi$ cannot be a subspace embedding for $U$.
We shall need the following lemmata.

The first lemma, inspired by %
Lemmata 8 and 9 in \cite{NN14}, shows that in a finite collection of vectors of length at most $1$ there always exists a small fraction of vector pairs whose inner product is not too small. %
We shall use this lemma to look for an inner product away from zero.
\begin{lemma}\label{lem: small inner product}
    Let $\eps\in(0, 1/9)$ and $\kappa = 3$. Suppose that $S$ is a finite subset of the unit $\ell_2$-ball and $u,v$ are independent samples from $\Unif(S)$. Then $\langle u,v\rangle\ge -\kappa\eps$ with probability at least $2\eps$.
\end{lemma}
\begin{proof}
    Note that
    \[
        0\le\left\|\sum_{u\in S}u\right\|_2^2 = \sum_{\substack{u,v\in S}}\langle u,v\rangle.
    \]
    Thus
    \[
        \E_{u,v}[\langle u,v\rangle]\ge 0.
    \]
    Since $\|u\|_2\le 1$ for all $u\in S$, it holds that $\langle u,v\rangle\le 1$.
    If $\Pr[\langle u,v\rangle \geq -\kappa \eps] \leq 2\eps$ for some $\kappa > 0$, then
    \[
        0\leq \E_{u,v}[\langle u,v\rangle] \leq 2\eps + (1-2\eps)(-\kappa\eps)
    \]
    which is true only if $\kappa < 2/(1-2\epsilon)$. %
    That is, when $\kappa = 3$ and $\eps\in (0,1/9)$, the inequality above will not hold and therefore $\Pr[\langle u,v\rangle \geq -\kappa \eps] > 2\eps$.
\end{proof}

The next lemma states that, 
if two columns of $\Pi V$ have a large inner product, then this large deviation implies the anti-concentration of $\Norm{\Pi U v}_2^2$ for some $v$, which refutes the existence of a good $\Pi$.

\begin{lemma}\label{lemma: inner product implies anticoncentration}
    Suppose that $|\langle A_{\ast,p},A_{\ast,q}\rangle| \ge \lambda\epsilon/\beta$ for some distinct columns $p,q$ of a matrix $A\in \mathbb{R}^{m\times d/\beta}$, where $\lambda>2$.
    Then there exists a unit vector $u\in\R^d$ such that with probability at least $1/4$ (over $W$)
    \[
        \Norm{AWu}_2^2 \notin [(1-\eps)^2, (1+\eps)^2].
    \]
\end{lemma}
\begin{proof}
    Let $p',q'$ be the columns of $W$ such that $W_{p,p'}\ne 0$ and $W_{q,q'}\ne 0$. First we assume that $p'\neq q'$.
    Let $u \triangleq (e_{p'}+e_{q'})/\sqrt{2}$, $S \triangleq \{i\in[d/\beta]: W_{i,p'}\ne 0 \text{ or } W_{i,q'}\ne 0\}$ and $\nu \triangleq \sum_{i\in S\setminus\{p,q\}} \sigma_i A_{\ast,i}$. Then
    \begin{equation}\label{eqn:coordinate_expansion}
        \Norm{AW(e_{p'}+e_{q'})}_2^2/\beta = \Norm{\sigma_p A_{\ast,p} + \sigma_q A_{\ast,q} + \nu}_2^2
        = \Norm{A_{\ast,p}}_2^2 + \Norm{A_{\ast,q}}_2^2 + \Norm{\nu}_2^2 + 2(X+Y+Z),
    \end{equation}
    where
    \[
    X\triangleq  \sigma_p\sigma_q \inner{A_{\ast,p}}{A_{\ast,q}}, \quad Y \triangleq  \sigma_p \inner{A_{\ast,p}}{\nu},\quad Z\triangleq  \sigma_q \inner{A_{\ast,q}}{\nu}.
    \]
    The assumption states that $|X|\geq \lambda\epsilon/\beta$ and thus $\max\{|X|,|Y|,|Z|\}\ge \lambda\epsilon/\beta$.
    We arbitrarily fix $\sigma_i$'s for all $i\in S\setminus\{p,q\}$ and then apply the following fact.
    \begin{fact}
        Suppose that $x_1,x_2,x_3\in\R$ satisfy $|x_1|\ge |x_2|\ge |x_3|$ and $|x_1|\ge a$. Let $\sigma_1,\sigma_2$ be independent Rademacher variables. Then $\Pr[\sigma_1x_1+\sigma_2x_2+\sigma_1\sigma_2x_3\ge a]\ge 1/4$ and $\Pr[\sigma_1x_1+\sigma_2x_2+\sigma_1\sigma_2x_3\le -a]\ge 1/4$.
    \end{fact}
    Note that $O \triangleq \Norm{A_{\ast,p}}_2^2 + \Norm{A_{\ast,q}}_2^2 + \Norm{\nu}_2^2$ is constant conditioned on $\{\sigma_i\}_{i\in S\setminus\{p,q\}}$. 
    We conclude that
    \begin{equation}\label{eqn:large tail}
        \Pr\left[\left.\Norm{AWu}_2^2 \ge \frac{O\beta}2 + \lambda\epsilon \ \right\vert \{\sigma_i\}_{i\in S\setminus\{p,q\}} \right]\ge \frac14\quad\text{and}\quad
        \Pr\left[\left.\Norm{AWu}_2^2 \le \frac{O\beta}2 - \lambda\epsilon \ \right\vert \{\sigma_i\}_{i\in S\setminus\{p,q\}} \right]\ge \frac14.
    \end{equation}
    Observe that $[(1-\eps)^2,(1+\eps)^2]$ is an interval of length $4\eps < 2\lambda\eps$, it follows that
    \[
    	\Pr\left\{ \left. \Norm{AWu}_2^2 \notin [(1-\eps)^2, (1+\eps)^2]\ \right\vert \{\sigma_i\}_{i\in S\setminus\{p,q\}}\right\} \geq 1/4.
    \]
    The claimed result then follows from the law of total probability. This completes the proof for the case $p'\neq q'$.
     
    Next we consider the case $p'=q'$. Let $u \triangleq e_{p'}$ and $S \triangleq \{i\in[d/\beta]: W_{i,p'}\ne 0\}$. Then the preceding argument goes through almost identically as $\Norm{AWe_{p'}}_2^2/\beta$ has exactly the same expansion as in \eqref{eqn:coordinate_expansion}. Instead of \eqref{eqn:large tail} we shall have
    \[
        \Pr\left[\left.\Norm{AWu}_2^2 \ge O\beta + 2\lambda\epsilon \ \right\vert \{\sigma_i\}_{i\in S\setminus\{p,q\}} \right]\ge \frac14 \quad\text{and}\quad
         \Pr\left[\left.\Norm{AWu}_2^2 \le O\beta - 2\lambda\epsilon \ \right\vert \{\sigma_i\}_{i\in S\setminus\{p,q\}} \right]\ge \frac14.
    \]
\end{proof}

\section{Lower Bound for $s=1$}\label{sec:1-sparse}

Our hard instance $\cD$ is $\cD_1$ with probability $1/2$ and $\cD_{8\eps}$ with probability $1/2$, where $\cD_\beta$ is defined in \Cref{def:D_beta}.

\begin{lemma}\label{claim:nonzeroes}
If $\Pi\in\mathbb{R}^{m\times n}$ is an $(\epsilon,\delta)$-subspace-embedding for $U\sim\cD$ with column sparsity $s=1$, then $(1-2\delta/d)$-fraction of the column vectors of $\Pi$ has $\ell_2$-norm of $1\pm\epsilon$, i.e. $(1-2\delta/d)$-fraction of the nonzero entries of $\Pi$ have absolute value $1\pm\epsilon$.
\end{lemma}
\begin{proof}
    By the law of total probability,
    \begin{multline*}
        1-\delta \le \Pr[\Pi\text{ is a subspace embedding for }U]\\
         = \frac{1}{2}\Pr[\Pi\text{ is a subspace embedding for }U|U\sim\cD_1] 
          + \frac{1}{2}\Pr[\Pi\text{ is a subspace embedding for }U|U\sim\cD_{8\epsilon}].
    \end{multline*}
    Thus we have $\Pr[\Pi\text{ is a subspace embedding for }U|U\sim\cD_1]\ge 1-2\delta$.
    
    Suppose that $\sigma$ fraction of the nonzero entries of $\Pi$ have the absolute value outside  $[1-\eps,1+\eps]$, then we have $\Pr[\|\Pi U e_i\|_2 \in [1-\eps,1+\eps] \mid U\sim\cD_1] = 1-\sigma$ for each $i\in[d]$. 
    Hence,
    \begin{align*}
        \Pr[\forall i\in[d],\|\Pi U e_i\|_2\in [1-\eps, 1+\eps] \mid U\sim\cD_1] = (1-\sigma)^d,
    \end{align*}
    and consequently, it must hold that $\sigma \leq 1 - (1-2\delta)^{1/d} \leq 2\delta/d$.
\end{proof}

For each $i\in [m]$, let $B_i$ denote the number of distinct $j\in[n]$ such that $|\Pi_{i,j}|\in [1-\eps,1+\eps]$ and $\Pi_{i,j}V_{j,l}\ne 0$ for some $l\in [d]$. Since $\Pi$ can be viewed as a hashing matrix of hashing $n$ items into $m$ buckets, $B_i$ can be viewed as the number of distinct dimensions which are hashed into the $i$-th bucket with small distortion (multiplied by $1\pm\eps$ or $-(1\pm\eps)$).
\begin{lemma}\label{claim: prob of disjoint}
    Let $\eps \in (0,1/8)$, $\delta\in (0,1/4)$ and $n\ge K d^2/(\epsilon^2\delta)$ for some absolute constant $K > 0$ large enough. If $\Pi\in\mathbb{R}^{m\times n}$ is an $(\epsilon,\delta)$-subspace-embedding for $U\sim\cD$ with column sparsity $s=1$, then conditioned on $U\sim\cD_{8\epsilon}$, with probability at most $2\delta/(1-4\delta)$, there exists an $i$ such that $B_i>1$.
\end{lemma}
\begin{proof}
    A similar argument to the beginning of the proof of \Cref{claim:nonzeroes} yields that 
    \[
    \Pr[\Pi\text{ is a subspace embedding for }U|U\sim\cD_{8\epsilon}]\ge 1-2\delta.
    \]
    We define an event $\cE$ to be that $B_i > 1$ for some $i\in [m]$. Let $\sigma \triangleq \Pr[\overline{\cE} \mid U\sim\cD_{8\epsilon}]$.
    Again by the law of total probability,
    \begin{align*}
         1-2\delta \le&  \Pr[\Pi\text{ is a subspace embedding for }U \mid U\sim\cD_{8\epsilon}]\\
        = & \Pr[\Pi\text{ is a subspace embedding for }U \mid U\sim\cD_{8\epsilon}, \overline{\cE}]\cdot\sigma\\
        & \quad +\Pr[\Pi\text{ is a subspace embedding for }U \mid U\sim\cD_{8\epsilon}, \cE]\cdot(1-\sigma).
    \end{align*}
    It follows that
    \[
    \Pr[\Pi\text{ is a subspace embedding for }U \mid U\sim\cD_{8\epsilon}, \cE]
     \ge \frac{1-2\delta-\sigma}{1-\sigma} =: 1-\sigma'.
    \]
    It suffices to show that $\Pi$ cannot be a subspace embedding for $U\sim\cD_{8\eps}$, conditioning on $\cE$ and an arbitrarily fixed $V$, with large probability.
    When $\cE$ happens, there exist $i\ne j$ such that $|\langle (\Pi V)_{\ast,i},(\Pi V)_{\ast,j}\rangle|\ge (1-\epsilon)^2 > 3/4$.
    Invoking \Cref{lemma: inner product implies anticoncentration}, we conclude that $\Pi$ is not a subspace embedding for $U$ with probability at least $(1-\sigma')/4$.

    Thus we have $(1-\sigma')/4\le \delta$, which implies that $1-\sigma\le 2\delta/(1-4\delta)$.
\end{proof}

\begin{theorem}\label{thm:1-sparse-lb}
    Let $\eps \in (0,1/8)$, $\delta\in (0,1/8)$, and $n\ge K d^2/(\epsilon^2\delta)$ for some absolute constant $K > 0$ large enough. If $\Pi\in\mathbb{R}^{m\times n}$ is an $(\epsilon,\delta)$-subspace-embedding for $U\sim\cD$ with column sparsity $s=1$, then $m =  \Omega(d^2/(\eps^2\delta))$.
\end{theorem}
\begin{proof}
It follows from \Cref{claim:nonzeroes}, together with linearity of expectation and Markov's inequality, that when conditioned on $U\sim\cD_{8\epsilon}$,  with probability at least $1-4\delta/d$, there are $d/(16\epsilon)$ entries of $U$ which are multiplied by $1\pm\epsilon$ or $-(1\pm\eps)$ in $\Pi U$.
By \Cref{claim: prob of disjoint} and a union bound, with probability at least $1-2\delta/(1-4\delta)-4\delta/d \geq 1-8\delta$, for any two distinct columns $i,j$ among the $d/(16\epsilon)$ columns of $V$, there does not exist $k\in [m]$ such that $(\Pi V)_{k,i}\ne 0$ and $(\Pi V)_{k,j}\ne 0$. 
The folklore lower bound for the birthday paradox immediately implies that the matrix $\Pi$ must have $\Omega(d^2/(\epsilon^2\delta))$ rows.
\end{proof}

\LinesNumbered
\begin{algorithm}[htp]
    \caption{Finding disjoint good column pairs}
    \label{algo: sampling pairs}
    Let $g$ be the number of good columns chosen by $V$ and let $S_1\gets[g]$\;
    Let $(C_1,C_2,\dots,C_g)\in[n]^g$ be the good columns among the $d$ columns chosen from $[n]$ by $V$ in the order they are sampled\;
    $G_1\gets G$\;
    $k\gets 1$\;
    \ForEach{$j\in[d/16]$}
    {
        \While{\True}
        {\label{alg:line:loop_internal_begins2}
            Compute $\phi_{k,c}\triangleq \Pr_{c'\sim \Unif(G_k)}[c'\match c]$ for all $c\in G_k$\;
            for all $l\in [m]$, $G_k^l\gets \{c\in G_k: |\Pi_{l,c}|\ge\sqrt{8\eps}\}$\;\label{alg:line:aux1.1}
            $\ell \gets \argmax_l |G_k^l|$\;
            $S'_k\gets\{i\in S_k: |\Pi_{\ell,C_i}|\ge\sqrt{8\eps}\}$\;\label{alg:line:aux1.2}\label{alg:line:loop_internal_ends2}
            \uIf{$\phi_{k,c} \le \eta/d$ for all $c\in G_k$}
            {
                $S'_k\gets\emptyset$\;\label{alg:line:loop_internal_begins}
                \Break\;
            }
            \lElseIf{$S'_k\ne\emptyset$}\Break
            output $(\ell,\perp)$\;
            $S_{k+1}\gets S_k$\;
            $G_{k+1}\gets G_k\setminus G_k^\ell$\;
            $k\gets k+1$\;
        }
        \uIf{$S'_k\ne\emptyset$}
        {
            \uIf{$|S'_k|\ge 2$}
            {
                sample two distinct $j',j''\in S'_k$\;
                output $(C_{j'},C_{j''})$\;\label{alg:line:output2}
                $S_{k+1}\gets S_k\setminus\{j',j''\}$\;
                $G_{k+1}\gets G_k$\;
            }
            \Else
            {
                output $(\ell,\perp)$\;
                $S_{k+1}\gets S_k\setminus S'_k$\;
                $G_{k+1}\gets G_k\setminus G_k^\ell$\;
            }
        }
        \uElseIf{$j\notin S_{k}$}
            {
                $S_{k+1}\gets S_{k}$\;
                $G_{k+1}\gets G_{k}$\;
                output $(\perp,\perp)$\;
            }
            \Else
            {
                $S'_k \gets \{i\in S_k\setminus \{j\}: C_i\match C_j\}$\;
                \uIf{$S'_k \neq \emptyset$}
                {
                    sample a $j'\sim \Unif(S_k')$\;
                    output $(C_{j'},C_j)$\; \label{alg:line:output}
                    $S_{k+1}\gets S_{k}\setminus \{j,j'\}$\;
                    $G_{k+1}\gets G_{k}$\;
                }
                \Else
                {
                    output $(\perp,C_j)$\;
                    $S_{k+1}\gets S_{k}\setminus\{j\}$\;
                    $G_{k+1}\gets G_k\setminus\{c\in G_k: c\match C_j\}$\; \label{alg:line:aux2}
                }
            } 
        $k\gets k+1$\;\label{alg:line:loop_internal_ends}
    }
\end{algorithm}

\section{Lower Bound for $s=1/(9\eps)$ with Abundance Assumption}
In this section we shall show an $\Omega(d^2)$ lower bound for all $\Pi$ with column sparsity $1/(9\eps)$ and an additional ``abundance assumption'' that $\Pi$ has many large entries in most columns (formalized as Assumption (ii) in Theorem~\ref{thm:with_avg_number}). We shall remove this assumption in the next section and obtain an $\Omega(\eps^{O(\delta)} d^2)$ lower bound.

For these lower bounds, we use the same distribution $\cD_\beta$ in Definition~\ref{def:D_beta} and consider a constant $\delta$. As explained after Definition~\ref{def:D_beta}, we pretend that $V$ has independent columns and the full column rank by assuming $n\geq Kd^2/(\beta^2\delta)$ for some $K$ large enough.

We say that a matrix entry is $\theta$-heavy if its absolute value is at least $\theta$.
We define the \emph{average number of $\theta$-heavy entries} of a matrix $A$ of $n$ columns as $\mathbb{E}_j\left[\left|\left\{i:|A_{i,j}|\ge \theta\right\}\right|\right]$, where $j\sim \Unif([n])$.
We say that two columns $i,j$ of $A$ share $k$ $\theta$-heavy rows if there exist $k$ distinct values $l\in[m]$ such that both $A_{l,i}$ and $A_{l,j}$ are $\theta$-heavy.

Below is the main theorem of the section.

\begin{theorem}\label{thm:with_avg_number}
    There exist absolute constants $\eps_0, \delta_0, K_0 > 0$ such that the following holds. For all $\eps\in(0,\eps_0)$, $\delta\in (0,\delta_0)$, $d\ge 1/\eps^2$ and $n\ge K_0 d^{2}/\delta$, 
    if
    \begin{enumerate*}[label=(\roman*)]
        \item 
    the column sparsity of $\Pi$ is at most $1/(9\eps)$, 
    \item 
    the average number of $\sqrt{8\eps}$-heavy entries of $\Pi$ is at least $1/(12\eps)$, and 
    \item 
    $\Pi$ is an $(\eps,\delta)$-subspace-embedding for $U\sim\cD_1$,
    \end{enumerate*}
 then $\Pi$ must have more than $d^2$ rows.
\end{theorem}
\begin{remark}\label{mark: tightness}
    The $d^2$ lower bound is tight up to a constant factor, demonstrated by the following example. Let $H$ be a Hadamard matrix of order $1/(8\eps)$. We define $\Pi$ to be the horizontal concatenation of copies of an $m \times m$ block diagonal matrix, where $m=O(d^2)$ and each diagonal block is $\sqrt{8\eps}H$. Clearly each column of $\Pi$ contains $1/(8\eps)$ entries of absolute value $\sqrt{8\eps}$. It can be easily verified that $\Pi$ is a $(0,\delta)$-subspace-embedding for $U\sim\cD_1$ and constant $\delta$.
\end{remark}

We shall prove Theorem~\ref{thm:with_avg_number} by contradiction. Assume that $\Pi$ has $m = d^2$ rows in the rest of this section. 

We say that two columns $i,j$ \emph{collide} with each other, denoted by $i\match j$, if they share at least one $\sqrt{8\eps}$-heavy row.
We call a column \emph{good} if it has at least $1/(16\eps)$ $\sqrt{8\eps}$-heavy entries and its $\ell_2$-norm is $1\pm\eps$.
Let $G\subseteq[n]$ be the set of indices of the good columns of $\Pi$. 
It follows from an averaging argument together with \cref{claim:nonzeroes} that at least $(1/3)$-fraction of the columns of $\Pi$ are good.

We are going to adopt the idea of the proof of Theorem~7 in \cite{NN14} to show that there is a pair of columns $(i,j)$ of $\Pi V$ such that $|\langle (\Pi V)_{\ast,i},(\Pi V)_{\ast,j}\rangle| = \Omega(\eps)$ with probability at least a constant.
To this end, we are going to find colliding column pairs and analyze their distributions.

\subsection{Finding colliding column pairs}
This subsection is devoted to analyzing Algorithm~\ref{algo: sampling pairs} and proving \Cref{lem: case 1 pairs,claim: disjoint pairs}, i.e. showing that Algorithm~\ref{algo: sampling pairs} finds many colliding pairs and they are well distributed.

For a sequence of column indices $(C_1,C_2,\dots,)\in[n]^*$, we say that two column pairs $(C_a,C_b),(C_c,C_d)\in [n]^2$ are disjoint if $a,b,c,d$ are distinct.
We select disjoint good column pairs among the good columns chosen by $V$ using the random process presented in Algorithm~\ref{algo: sampling pairs}, in which we set $\eta\triangleq 3$.
Our goal is to show that  with constant probability, there is a column pair $(C_i,C_j)$, found by Algorithm~\ref{algo: sampling pairs}, such that $|\langle \Pi_{\ast,C_i},\Pi_{\ast,C_j}\rangle|=\Omega(\eps)$.

Suppose that the output of Algorithm~\ref{algo: sampling pairs} are $Y_1,\dots,Y_{d/16}$. Each $Y_i$ is a random pair over $(\{\perp\}\cup[n])\times(\{\perp\}\cup[n])$. We have the following lemma on the guarantee of Algorithm~\ref{algo: sampling pairs}. %
\begin{lemma}\label{lem:sets of good columns}
    The set sequences $\{G_k\}$ and $\{S_k\}$ are both decreasing and nonempty. 
    Furthermore, %
    conditioned on $Y_1,\dots,Y_{k-1}$, the set $G_k$ is determined and the random variables $\{C_i\}_{i\in S_k}$ are independent and uniformly distributed over $G_k$.
\end{lemma}
\begin{proof}
    It is clear from the algorithm that $G_k$ is determined by $Y_1,\dots,Y_{k-1}$ and both $\{G_k\}$ and $\{S_k\}$ are non-increasing and nonempty.
    It remains to show that $\{C_i\}_{i\in S_k}$ are i.i.d.\ $\Unif(G_k)$, which we shall prove by induction on $k$.
    
    The base case is $k=1$.
    Indeed, all the $g$ $C_i$'s are independent and uniformly distributed over $G$. Since $G_1 = G$, the base case is verified.

    For the induction step, we fix a $k$ such that $1 \le k$. We also fix $Y_1,\dots,Y_{k-1}$. 
    
    If there exists $c\in G_k$ such that $\phi_{k,c}>\eta/d$ and $Y_k=(\cdot,\perp)$, then we know that $S_{k+1} = S_k\setminus S'_k$ and $G_{k+1} = G_k\setminus G_k^\ell$, where $G_k^\ell$ and $S'_k$ are as defined in Lines~\ref{alg:line:aux1.1} and~\ref{alg:line:aux1.2}.
    For each $i\in S_{k+1}=S_k\setminus S'_k$, we know that $\Pi_{\ell,C_i}$ is not heavy and thus $C_i\notin G_{k}^\ell$. 
    Since $C_i$ is uniform on $G_k$ given $Y_1,\dots,Y_{k-1}$, we know that $C_i$ is uniform on $G_k\setminus G_k^\ell$ when further conditioned on that $C_i\not\in G_k^\ell$. 
    The induction hypothesis also implies that $\{C_i\}_{i\in S_{k+1}}$ are independent.

    If there exists $c\in G_k$ such that $\phi_{k,c}>\eta/d$ and $Y_k\neq (\cdot,\perp)$, we know that $G_{k+1} = G_k$ and $S_{k+1}\subset S_k$. 
    Since $\{C_i\}_{i\in S_{k+1}}$ is a subset of $\{C_i\}_{i\in S_{k}}$, which are i.i.d.\ $\Unif(G_k)$ by the induction hypothesis, it is clear that $\{C_i\}_{i\in S_{k+1}}$ are i.i.d.\ $\Unif(G_k)$.

    Now we assume that $\phi_{k,c}\leq \eta/d$ for all $c\in G_k$.
    When $Y_k=(\perp,\perp)$, then we know $j\notin S_k$. 
    Since $G_{k+1} = G_k$ and $S_{k+1} = S_k$, it follows immediately from the induction hypothesis that $\{C_i\}_{i\in S_{k+1}}$ are i.i.d.\ $\Unif(G_{k+1})$.
    When $Y_k\in[n]\times [n]$, we know that $G_{k+1}=G_k$ and $S_{k+1}\subset S_k$, thus it follows from the induction hypothesis that $\{C_i\}_{i\in S_{k+1}}$ are i.i.d.\ $\Unif(G_{k+1})$. 
    When $Y_k \in \{\perp\}\times[n]$, for each $i\in S_{k+1}$, it must hold that $C_i$ does not collide with $C_j$ and thus $C_i\not\in \{c\in G_k: c\match C_j\}$ by the Line~\ref{alg:line:aux2}. 
    Since $C_i$ is uniform on $G_k$ given $Y_1,\dots,Y_{k-1}$, we know that $C_i$ is uniform on $G_{k+1}$ when further conditioned on that $C_i\in G_{k+1}$. 
    The induction hypothesis also implies that $\{C_i\}_{i\in S_{k+1}}$ are independent.
    
    The proof of the inductive step is now complete.
\end{proof}

Let $\cE$ denote the event that at least $7d/24$ columns of $\Pi V$ are good. By a Chernoff bound, we know that $\Pr(\cE) \geq 1-\exp(-\Omega(d))$.
When $\cE$ happens, the set $S_1$ in Algorithm~\ref{algo: sampling pairs} contains at least $7d/24$ good column indices with high probability in $d$. 

For each $k$ and each $c\in G_k$ (where $G_k$ is as guaranteed by Lemma~\ref{lem:sets of good columns}), we define
\begin{equation}\label{eqn:phi}
\phi_{k,c} \triangleq \Pr_{c'\sim \Unif(G_k)}[c'\match c].
\end{equation}
Observe that the while-loop never shrinks $S_k$ and the for-loop remove at least one and at most two elements from $S_k$ during each iteration. It is clear that $|S_{k}\setminus\{j\}|\ge (7d/24) - 2\cdot(d/16) = d/6$ for all $j \leq k$. Furthermore, it is easy to see that the while-loop will also be broken. We shall discuss two cases based on the breaking conditions of the while-loop.

\begin{lemma}\label{lem: case 1 pairs}
    Assume that $\cE$ happens. 
    For each $j\in[d/16]$, if the while-loop is broken by the event $S'_k\ne\emptyset$, then with probability at least $\delta'\eps$, Algorithm~\ref{algo: sampling pairs} outputs a column pair $(C_{j'},C_{j''})\sim \Unif(G_k^\ell\times G_k^\ell)$, where $\delta' > 0$ is an absolute constant. %
\end{lemma}
\begin{proof}
    When the while-loop is broken by event $S'_k\ne\emptyset$, it must happen that $\phi_{k,c}>\eta/d$ for some $c\in G_k$. 
    Let $M \triangleq \{l\in [m]: |\Pi_{l,c}|\ge \sqrt{8\eps}\}$.
    Note that
    $
        \{c'\in G_k:c'\match c\} = \bigcup_{l\in M} G_k^l
    $.
    The choice of $\ell$ implies that $|G_k^\ell|\ge |\{c'\in G_k:c'\match c\}|/|M_c|$. %
    Since the column sparsity of $\Pi$ is $s$, $|M_c|\le s=1/(9\eps)$ and so $|G_k^\ell|\ge (\phi_{k,c}\cdot |G_k|)/|M_c|\geq \eta\eps/(9d)\cdot|G_k|$. Let $p\triangleq |G_k^\ell|/|G_k|$, so $p\geq \eta\eps/(9d)$.

    Observe that at most two elements are removed from $S_k$ during each iteration of the for-loop.
    Since $\cE$ happens, it always holds that $|S_k|\geq d/6$.
    By Lemma~\ref{lem:sets of good columns}, the elements in $S_k'$ are sampled independently from $G_k$ with sampling probability $p$. Hence,
    \begin{align*}
    \Pr[|S'_k|\ge2\mid S'_k\ne\emptyset] = 1 - \frac{\Pr[|S'_k| = 1]}{\Pr[S'_k\ne\emptyset]} &= 1 - \frac{|S_k| p(1-p)^{|S_k|-1}}{1-(1-p)^{|S_k|}}\\
    &\geq 1 - \frac{|S_k| p(1-p)^{|S_k|-1}}{ p |S_k|}\\
    &= 1 - (1-p)^{|S_k|-1}.
    \end{align*}
    Observe that the rightmost side above is increasing w.r.t.\ $p$. When $p\cdot |S_k| \leq 1$,
    \[
    \Pr[|S'_k|\ge2\mid S'_k\ne\emptyset] \geq \left(1-\frac{1}{e}\right)p(|S_k|-1) \geq \left(1-\frac{1}{e}\right)\frac{\eta}{54}\eps.
    \]
    Therefore, it always hold that $\Pr[|S'_k|\ge2\mid S'_k\ne\emptyset] \geq \delta'\cdot \eps$, where $\delta' = (1-1/e)\eta/54$.
\end{proof}

Now we assume that the while-loop is broken by the event that ``$\phi_{k,c} \leq \eta/d$ for all $c\in G_k$''. 
We further define $T_k \triangleq \{(i,i')\in G_k\times G_k: i\match i'\}$, i.e., the set of good column pairs in $G_k$ which collide with each other. 
Let $\cD'_k$ denote the distribution of the output $(C_{j'},C_j)$ in Line~\ref{alg:line:output} of Algorithm~\ref{algo: sampling pairs}. 
Let $\Delta_k$ be a random variable representing the conditional expected number of $\sqrt{8\eps}$-heavy rows shared by two random good columns from $T_k$ given $Y_1,\dots,Y_{k-1}$. 
Note that $\forall c\in G_k,(c,c)\in T_k$, hence $T_k\ne\emptyset$, and thus $\Delta_k$ is well-defined.

\begin{lemma}\label{claim: disjoint pairs}
    Assume that $\cE$ happens. For each $j\in[d/16]$, if the while-loop is broken by event ``$\phi_{k,c} \leq \eta/d$ for all $c\in G_k$'' and if $j\in S_k$, then with probability at least $d/(K\eps^2  m\Delta_k)$, Algorithm~\ref{algo: sampling pairs} outputs a column pair $(C_{j'},C_j)$, where $K > 0$ is an absolute constant. Furthermore, the probability mass function of $\cD'_k$ differs from that of $\Unif(T_k)$ by at most a factor of an absolute constant.
\end{lemma}

\begin{proof}
    We claim that %
    \begin{equation}\label{EQN:COLLISION OUTSIDE G'_J}
    \sum_{c\in G_k} \frac{\phi_{k,c}}{|G_k|}  = \Pr_{c,c'\sim \Unif(G_k)}[c'\match c]\ge \frac{1}{256\eps^2 m\Delta_k}.
    \end{equation}
    Indeed, let $X$ denote the number of $\sqrt{8\eps}$-heavy rows shared by two random good columns $i,i'\sim\Unif(G_{k})$.
    Note that 
    \begin{align*}
       \E[X] &= \sum_{l\in[m]}\Pr[\Pi_{l,i},\Pi_{l,i'}\text{ are }\sqrt{8\eps}\text{-heavy}]\\
       &= \sum_{l\in[m]}\Pr[\Pi_{l,i}\text{ is }\sqrt{8\eps}\text{-heavy}]^2\\
       &\ge \frac{1}{m}\left(\sum_{l\in [m]} \Pr[\Pi_{l,i}\text{ is }\sqrt{8\eps}\text{-heavy}]\right)^2 \quad \text{(Cauchy-Schwarz)}\\
       &\ge \frac{1}{m}\cdot\left(\frac{1}{16\eps}\right)^2 = \frac{1}{256\eps^2 m}. 
    \end{align*}
    Since $X\geq 0$, applying the law of total expectation gives that 
    \[
        \E[X] = \E[X|X=0]\cdot\Pr[X=0]+\E[X|X>0]\cdot\Pr[X>0] = \Delta_k\cdot\Pr[X>0],
    \]
    which implies that $\Pr[X>0]= \E[X] / \Delta_k \ge 1/(256\eps^2 m\Delta_k)$. This establishes~\eqref{EQN:COLLISION OUTSIDE G'_J}.

    Observe that at most two elements are removed from $S_k$ during each iteration of the for-loop.
    Since $\cE$ happens, it always holds that $|S_k|>d/6$.
    Thus, $C_j$ collides with some $C_{j'}\in G_j$, where $j'\in S_k\setminus\{j\}$, with probability at least
    \begin{align*}
     \sum_{c\in G_k} \Pr[C_j=c] \left(1 - (1 - \phi_{k,c})^{\frac{d}{6}}\right) 
    \geq \sum_{c\in G_k} \frac{1}{|G_k|} \cdot \left(1-\frac{1}{e}\right)\cdot \phi_{k,c}\cdot\frac{d}{6} \geq \frac{d}{K}\cdot\frac{1}{256\eps^2 m\Delta_k}.
    \end{align*}
    for some absolute constants $K>0$.
    Here, the first inequality follows from the fact that $\phi_{k,c}\cdot d/6 \leq \eta/6 \leq 1$ and the second inequality from~\eqref{EQN:COLLISION OUTSIDE G'_J}.     
    
    Next we analyze the distribution of the colliding pair $(C_{j'},C_j)$. Given that there is such an output, the distribution of $(C_{j'},C_j)$ can be obtained by first sampling $C_j = c\in G_k$ with probability $p_{k,c} \triangleq \phi'_{k,c}/\sum_{c\in G_k} \phi'_{k,c}$, where 
    \[
    \phi'_{k,c} \triangleq \Pr%
    [\exists c'\in \{C_i\}_{i\in S_k\setminus\{j\}}, c'\match c] = 1 - (1-\phi_{k,c})^{|S_k|-1},
    \]
    and then independently sampling $j'\sim \Unif(S_k')$. Since we are conditioning on that the algorithm outputs a column pair, the second step is equivalent to choosing $C_{j'}\sim \Unif(M_{C_j})$, where $M_c\triangleq \{c'\in G_k:c'\match c\}$. 
    
    Now we examine $(C',C)\sim\Unif(T_k)$. Then $\Unif(T_k)$ can be obtained from first sampling $C = c\in G_k$ with probability $q_{k,c} \triangleq \phi_{k,c}/\sum_{c\in G_k}\phi_{k,c}$ and then independently sampling $C'\sim \Unif(M_c)$. 
    
    The result follows from the fact that 
    \[
        p_{k,c} = \frac{\phi_{k,c}'}{\sum_{c\in G_k} \phi_{k,c}'} = \Theta\left( \frac{\phi_{k,c}}{\sum_{c\in G_k}\phi_{k,c}} \right) = \Theta(q_{k,c}),\quad \forall c\in G_k. \qedhere
    \]
\end{proof}

\subsection{Looking for large inner product}
In this subsection, we shall show that (informally)
\begin{enumerate}
    \item (Lemma~\ref{lem: large inner product}) every column pair $(i,j)$, found by Algorithm~\ref{algo: sampling pairs} in Line~\ref{alg:line:output}, satisfies that $|\langle \Pi_{\ast,i},\Pi_{\ast,j}\rangle|=\Omega(\eps)$ with probability $\Omega(d/m)$;
    \item (Corollary~\ref{corollary: large inner product}) Algorithm~\ref{algo: sampling pairs} finds a column pair whose inner product is $\Omega(\eps)$ with probability $\Omega(d^2/m)$.
\end{enumerate}

Recall that the column sparsity is at most $s = 1/(9\eps)$. We further define for each $x\in[s]$ 
\[
\left\{
\begin{aligned}
Q_{x,k} &\triangleq \{(i,i')\in T_k: \text{$i$ and $i'$ share exactly $x$ $\sqrt{8\eps}$-heavy rows}\}, \\
P_{x,k} &\triangleq \{(i,i')\in Q_{x,k}: \langle \Pi_{\ast,i}, \Pi_{\ast,i'} \rangle\geq (8-\kappa)\eps \}.
\end{aligned}
\right.
\]
Let $q_{x,k} = |Q_{x,k}|/|T_k|$ and $p_{x,k} = |P_{x,k}|/|T_k|$. Note that $\{q_{x,k}\}_x$ forms a probability distribution on $[s]$ and $\sum_x x\cdot q_{x,k}=\Delta_k$.

To analyze $\cD'_k$, which is close to $\Unif(T_k)$, we consider an analogous distribution $\cD''_k$.
Instead of sampling a good column pair uniformly at random over $T_k$, $\cD''_k$ is obtained by sampling a good column pair $(i,i')$ with probability proportional to the number of $\sqrt{8\eps}$-heavy rows shared by columns $i$ and $i'$.

We need the following observation.
\begin{lemma}\label{claim: large inner product}
Let $\eps\in(0,1/9)$. 
    Suppose that a matrix $A$ has a row $l$ such that $S \triangleq \{i: |A_{l,i}|\geq \theta\}$ is nonempty. Further suppose that $\Norm{A_{\ast,i}}_2^2\leq 1 + \theta^2$ for all $i\in S$. Let $u,v\sim \Unif(S)$ be independently sampled.
    Then $\langle A_{\ast,u},A_{\ast,v}\rangle \ge \theta^2 - \kappa\eps$ with probability at least $\eps/2$, where $\kappa$ is as in \Cref{lem: small inner product}.
\end{lemma}
\begin{proof}
    We partition $S$ into two groups $S^- = \{i: A_{l,i} < 0\}$ and $S^+ = \{i: A_{l,i} > 0\}$.
    Without loss of generality, we assume that $|S^-|\ge|S^+|$.
    With probability at least $1/4$, we have $u,v\in S^-$. Condition on this event. It is clear that we have $A_{l,u}A_{l,v}\ge \theta^2$ and $u,v$ are independent uniform samples over $S^-$. %
    
    Let $S'$ be the set of column vectors consisting of all the column vectors in $S^-$ with the $l$-th entry removed. Applying \Cref{lem: small inner product} to $S'$ yields that $\langle A_{\ast,u},A_{\ast,v}\rangle - A_{l,u}A_{l,v} \ge - \kappa\eps$ with probability larger than $2\eps$. 
    Unconditioning on $u,v\in S^-$, the claim follows.
\end{proof}

The following is an immediate corollary of the preceding lemma.
\begin{corollary}\label{lem:key_ratio}
It holds that 
$
\big(\!\sum_{x=1}^{s} x\cdot p_{x,k}\big)\big/\big(\!\sum_{x=1}^{s} x \cdot q_{x,k}\big) \geq \eps/2.
$
\end{corollary}
\begin{proof}
Note that $\cD''_k$ can be rephrased as samples some row $l$, then sample two good columns $(i,i')$ such that $|\Pi_{l,i}|,|\Pi_{l,i'}|\ge\sqrt{8\eps}$. \Cref{claim: large inner product} implies the advertized result.
\end{proof}

Let $\hat p_k \triangleq \sum_x p_{x,k}$, then $\hat p_k$ has the combinatorial interpretation that
    \[
        \hat p_k = \Pr_{(i,i')\sim\Unif(T_k)}\left[\langle \Pi_{\ast,i},\Pi_{\ast,i'}\rangle \ge (8-\kappa)\eps\right].
    \]
For the true distribution $\cD_k'$, define correspondingly
\[
    \widetilde{p}_k = \Pr_{(i,i')\sim\cD_k'}\left[\langle \Pi_{\ast,i},\Pi_{\ast,i'}\rangle \ge (8-\kappa)\eps\right].
\]
Recall that $\cD'_k$ is different from $\Unif(T_k)$ by a constant factor, we have $\widetilde{p}_k = \Theta(\hat{p}_k)$.
We shall show that 
if $j\in S_{k}$ and $\phi_{k,c}\le \eta/d$ for all $c\in G_k$, with probability at least $\Omega(d/m)$, Algorithm~\ref{algo: sampling pairs} in Line~\ref{alg:line:output} outputs a pair $(C_{j'},C_j)$ such that  $\langle \Pi_{\ast,C_{j'}},\Pi_{\ast,C_j}\rangle \ge (8-\kappa)\eps$.

\begin{lemma}\label{lem: large inner product}
    There exists an absolute constant $\delta' > 0$ such that the following holds. Assume that $\cE$ happens.
    Suppose that $\eps\in(0,1/9)$, the while-loop is broken by event ``$\phi_{k,c} \leq \eta/d$ for all $c\in G_k$'', and $j\in S_k$.
    With probability at least $\delta'd/m$, Algorithm~\ref{algo: sampling pairs} in Line~\ref{alg:line:output} outputs a pair $(C_{j'},C_j)$ such that  $\langle \Pi_{\ast,C_{j'}},\Pi_{\ast,C_j}\rangle \ge (8-\kappa)\eps$,
    where $\kappa$ is as in \Cref{lem: small inner product}.
\end{lemma}
\begin{proof}
    Since $\Delta_k\le s=1/(9\eps)$, Algorithm~\ref{algo: sampling pairs} in Line~\ref{alg:line:output} outputs a pair $(C_{j'},C_j)$ with probability at least $9d/(K\eps m)$ by \Cref{claim: disjoint pairs}. When $\hat p_k\ge \eps/12$, we have that $\widetilde{p}_k = \Theta(\hat p_k) = \Omega(\eps)$ and the desired result follows.

    Next we assume that $\hat p_k\le \eps/12$.
    Let $\eps' \!\triangleq \eps/2$, $w_k \!\triangleq\! (\sum_x x\cdot p_{x,k})/\hat p_k$ and $\Delta'_k \triangleq \sum_x x(q_{x,k}-p_{x,k})$. 
    It is clear that $w_k\leq (\sum_x s\cdot p_{x,k})/\hat p_k = s$.
    Note that $\Delta_k=\Delta'_k + w_k\cdot\hat p_k = \sum_x x\cdot q_{x,k} \geq \sum_x q_{x,k} = 1$, we know that $\Delta'_k \geq 1-s\cdot\hat p_k\geq 107/108\ge s\cdot \hat p_k\ge w_k\cdot\hat p_k $, so $\Delta'_k\ge \Delta_k/2$.
    The inequality in \Cref{lem:key_ratio} can be written as
    \[
      \frac{w_k\hat p_k}{\Delta'_k + w_k\hat p_k}  \geq \eps'.
    \]
    Thus,
    \[
        \hat p_k\geq \frac{\eps' \Delta'_k}{(1-\eps')w_k}\geq \frac{\eps' \Delta'_k}{s} = \frac{(\eps/2)\Delta'_k}{1/(9\eps)} = \frac{9\eps^2 \Delta'_k}{2}\ge \frac{9\eps^2 \Delta_k}{4}.
    \]
    The conclusion then follows from \Cref{claim: disjoint pairs} and the fact that $\widetilde{p}_k = \Theta(\hat p_k)$.
\end{proof}

\begin{corollary}\label{corollary: large inner product}
    There exist absolute constants $\delta'', K > 0$ such the following holds. Suppose that $d\ge 1/\eps^2$ and $\eps\in(0,1/9)$. Assume that $\cE$ happens.
    Then, with probability at least $\delta''\cdot\min\{d^2/m,1\}$, Algorithm~\ref{algo: sampling pairs} outputs a pair $(C_i,C_{i'})$ such that $\langle \Pi_{\ast,C_i},\Pi_{\ast,C_{i'}}\rangle \ge (8-\kappa)\eps$, where $\kappa$ is as in \Cref{lem: small inner product}.
\end{corollary}
\begin{proof} 
    Observe that the while-loop never shrinks $S_k$ and the for-loop always removes one or two elements from $S_k$ during each iteration.
    We claim that, assuming $\cE$ happens, either 
    \begin{enumerate*}[label=(\roman*)]
        \item 
    ``the while-loop is broken by event $S'_k\ne\emptyset$''
    happens for at least $d/96$ times, or
        \item
    ``the while-loop is broken by event $\forall c\in G_k,\phi_{k,c} \leq \eta/d$'' and ``$j\in S_k$''
    happens simultaneously for at least $d/64$ times.
    \end{enumerate*}
    We analyze the two cases separately.
    
    In case (i), by \cref{lem: case 1 pairs}, Line~\ref{alg:line:output2} of Algorithm~\ref{algo: sampling pairs} outputs a distinct column pair with probability $\Omega(\eps)$. Let $\cF$ denote the event that Line~\ref{alg:line:output2} outputs $\Omega(\eps d)$ distinct pairs over the entire execution, then $\Pr[\overline{\cF}]\leq \exp(-\Omega(\eps d))$ by a Chernoff bound.
    By \cref{claim: large inner product}, each pair has a desired large inner product with probability $\eps/2$. 
    Hence, when $\cF$ happens, since $d\ge 1/\eps^2$, one of the pairs has a desired large inner product with probability at least $1-(1-\eps/2)^{\Omega(\eps d)}\geq 1-e^{-c\eps^2 d} \geq 1-e^{-c}$ for some constant $c > 0$. One can choose $\eps_0$ to be small enough such that $\Pr[\overline{\cF}]\leq \exp(-\Omega(1/\eps))\leq (1-e^{-c})/2$ so that the overall success probability in case (i) is at least $(1-e^{-c})/2$.
    
    For case (ii), immediately by \cref{lem: large inner product} we obtain that the desired probability is lower bounded by 
    \[
    1-\left(1-\frac{\delta'd}{m}\right)^{d/64}\geq \frac{\delta'' d^2}{m}
    \] for some absolute constant $\delta'' > 0$.
\end{proof}

\subsection{Putting everything together}
Now we are ready to complete the proof of \Cref{thm:with_avg_number}.

\begin{proof}[Proof of \Cref{thm:with_avg_number}]
Since $n\ge K_0d^2/\delta$, $V$ does not have the full column rank with probability at most $\delta/(2K_0)\le \delta''$ for large enough constant $K_0>0$, where $\delta''$ is as in \Cref{corollary: large inner product}.
Suppose that $d$ is at least a sufficiently large constant $d_0$ such that $\Pr\big[\overline{\cE}\big]  \leq \exp(-\Omega(d)) < \delta''$.

Let $\delta_0 = \delta'' / 5$ and $\eps_0 = \min\{1/\sqrt{d_0}, 1/9\}$.
When conditioned on $\cE$, it follows from \Cref{corollary: large inner product} that with probability at least $\delta''$ there is a pair $(i,j)$ such that $\langle (\Pi V)_{\ast,i},(\Pi V)_{\ast,j}\rangle \ge (8-\kappa)\eps$.
Noting that $8 - \kappa > 2$, we invoke Lemma~\ref{lemma: inner product implies anticoncentration} and conclude that $\Pi$ cannot be an $(\eps,\delta)$-subspace-embedding for $U\sim \cD_1$, whenever $\delta < \delta_0\le (\delta''/4)\cdot(1-2\delta'')$.
Therefore, $\Pi$ must have more than $d^2$ rows.
\end{proof}

\section{Removing the Abundance Assumption}

In this section, we remove the ``abundance assumption'', i.e. Assumption (ii) in Theorem~\ref{thm:with_avg_number}, and prove the \cref{thm:without_avg_number}.

Consider a distribution $\widetilde{\cD}$ over $\R^{n\times d}$: with probability $1/2$, it is $\cD_1$; with probability $1/2$, it is $\cD_{2^{-\ell}}$ for an uniformly chosen $\ell\in[L]$, where $L \triangleq \log(1/\eps)-3$.

\begin{theorem}\label{thm:without_avg_number}
There exist absolute constants $\eps_0, \delta_0, c_0, K_0, K_1 > 0$ such that the following holds.
For all $\eps \in (0,\eps_0)$, $\delta \in (0,\delta_0)$, $d\ge 1/\eps^2$ and $n\ge K_0 d^{2}/(\eps^2\delta)$, 
if
\begin{enumerate*}[label=(\roman*)]
    \item the column sparsity of $\Pi$ is at most $1/(9\eps)$, and
    \item $\Pi$ is an $(\eps,\delta)$-subspace-embedding for $U\sim\widetilde{\cD}$,
\end{enumerate*}
then $\Pi$ must have more than $c_0 \log^{-4}(1/\eps)\epsilon^{K_1\delta} d^2$ rows.
\end{theorem}

The rest of the section is devoted to the proof of \Cref{thm:without_avg_number}. Let 
\[
    \delta'\triangleq\frac{\log\log(1/\epsilon^{72})}{\log(1/\epsilon)},
\]
then $4\epsilon^{\delta'}\log(1/\epsilon)\le 1/18$.

We start by assuming there is a matrix $\Pi\in\mathbb{R}^{m\times n}$ which is an $(\eps,\delta)$-subspace-embedding of $U\sim\widetilde{\cD}$ for $m = \eps^{4\delta'+{K_1\delta}}d^2$.

By the law of total probability and the averaging principle, $\Pi$ is an $(\eps,2\delta)$-subspace-embedding for $U\sim\cD_1$, and by Markov's inequality, for $(1-\gamma)$-fraction of $\ell'\in[L]$, $\Pi$ is an $(\eps,2\delta/\gamma)$-subspace-embedding for $U\sim\cD_{2^{-\ell'}}$, where $\gamma\triangleq K_1\delta/2 < 1$.
Hence, for all $\ell\in [L]$, there exists an 
\[
   \ell'\in[\max\{0,(1-\gamma)L-\ell\},L-\ell] \text{ such that }2^{-\ell-\ell'}\in[ 2^{-L} , (2^{-L})^{1-\gamma}]
\]  and $\Pi$ is an $(\eps,2\delta/\gamma)$-subspace-embedding for $U\sim\cD_{2^{-\ell'}}$.

Let $G_0\subseteq[n]$ denote the set of the column indices $i\in[n]$ such that $\|\Pi_{\ast,i}\|_2 =1\pm\eps$.
By \Cref{claim:nonzeroes}, $|G_0|\ge (1-2\delta/d)n$.
Let $\Pi'$ be the submatrix of $\Pi$ which consists of all the columns in $G_0$.

Note that the following lemma implies that the average squared $\ell_2$-norm of the columns of $\Pi'$ is at most $4\eps^{\delta'}\log(1/\eps)+s(8\eps)\le 17/18 < (1 - \eps)^2$.
Therefore, it suffices to prove the following lemma to refute $\Pi$.
\begin{lemma}\label{lem:key_general_sparsity}
    Suppose that $\eps\in(0,1/9)$, $d\ge 1/\eps^{2}$, $n\ge K_0 d^2/(\eps^2\delta)$ and $\Pi$ is an $(\eps,\delta)$-subspace-embedding for $U\sim\widetilde{\cD}$.
    For every $\ell\in [L]$, the average number of $\sqrt{2^{-\ell}}$-heavy entries of $\Pi'$ is at most $\eps^{\delta'}2^{\ell}$.
\end{lemma}
\LinesNotNumbered
\begin{algorithm}[tp]
    \caption{Finding disjoint good column pairs}
    \label{algo: sampling pairs 2}
    Let $g$ be the number of good columns chosen by $V$ and let $S_1\gets[g]$\;
    Let $(C_1,C_2,\dots,C_g)\in[n]^g$ be the good columns among the $\epsilon^{\delta'} d\cdot 2^{\ell'}$ columns chosen from $[n]$ by $V$ in the order they are sampled\;
    $G_1\gets G$\;
    $k\gets 1$\;
    \ForEach{$j\in[\epsilon^{\delta'} d\cdot 2^{\ell'}/16]$}
    {
            $\vdots$\\
            (the same as Lines \ref{alg:line:loop_internal_begins2}--\ref{alg:line:loop_internal_ends2} of Algorithm~\ref{algo: sampling pairs})\\
            $\vdots$\\
            \uIf{$\phi_{k,c} \le \eta/(\eps^{\delta'}d2^{\ell'})$ for all $c\in G_k$}{}
            $\vdots$\\
            (the same as Lines \ref{alg:line:loop_internal_begins}--\ref{alg:line:loop_internal_ends} of Algorithm~\ref{algo: sampling pairs})\\
            $\vdots$\\
    }
\end{algorithm}
\begin{proof}[Proof sketch]
    We prove the lemma by contradiction.
    Fix an arbitrary $\ell\in [L]$ and suppose that the average number of $\sqrt{2^{-\ell}}$-heavy entries of $\Pi'$ is at least $\eps^{\delta'}2^{\ell}$.
    By the arguments preceding \Cref{lem:key_general_sparsity}, $\Pi$ is an $(\eps,2\delta/\gamma)$-subspace-embedding for $U\sim\cD_{2^{-\ell'}}$.
    
    We say that two columns $i,j$ \emph{collide} with each other, denoted by $i\match j$, if they share at least one $\sqrt{2^{-\ell}}$-heavy row.
    We call a column \emph{good} if it has at least $\eps^{\delta'} 2^{\ell}/3$ $\sqrt{2^{-\ell}}$-heavy entries and its $\ell_2$-norm is $1\pm\epsilon$.
    
    Recall that every column of $\Pi'$ has the $\ell_2$-norm within $[1-\eps,1+\eps]$. Hence, in each column of $\Pi'$, the number of $\sqrt{2^{-\ell}}$-heavy entries is at most $2^\ell(1+\eps)^2$.
    We claim that at least $(\eps^{\delta'}/2)$-fraction of the columns of $\Pi$ are good. Indeed, let $\theta$ be the fraction of good columns of $\Pi'$, then we have
    \[
    \eps^{\delta'} 2^\ell \leq \theta 2^\ell (1+\eps)^2 + (1-\theta)\frac{\eps^{\delta'} 2^\ell}{3}
    \leq \frac{100}{81}\theta 2^\ell + (1-\theta)\frac{\eps^{\delta'} 2^\ell}{3},
    \]
    whence we can solve that $\theta \geq \eps^{\delta'}\cdot (2/3)/(100/81-\epsilon^{\delta'}/3) \geq \eps^{\delta'}/2$.
    
    Recall that $V$ has $d' \triangleq d2^{\ell'}$ columns.
    Since $d\ge 1/\eps^2$ and $2^{\ell'}\ge 1$, it holds that $\eps^{\delta'} d' \geq 1/(\eps^2\polylog(1/\eps))$.
    By a Chernoff bound, with probability at least $1-\exp(-1/\eps^{2-o(1)})$, there exist $\eps^{\delta'} d'/4$ good columns in $\Pi V$. 
    
    We follow the proof of \Cref{lem: case 1 pairs} to show that if the while-loop is broken by event $S'_k\ne\emptyset$, then Algorithm~\ref{algo: sampling pairs 2} in Line~\ref{alg:line:output2} outputs a pair $(C_{j'},C_j)$ with probability $\Omega(\eps)$.
    
    We follow the proof of \Cref{claim: disjoint pairs} to show that if the while-loop is broken by event ``$\phi_{k,c} \leq \eta/d$ for all $c\in G_k$'' and if $j\in S_k$, then Algorithm~\ref{algo: sampling pairs 2} in Line~\ref{alg:line:output} outputs a pair $(C_{j'},C_j)$ with probability $\Omega(\epsilon^{3{\delta'}}d\cdot 2^{2\ell+\ell'}/(m\Delta_{\ell,j}))$.

    Then, following the proof of \Cref{corollary: large inner product}, we conclude that Algorithm~\ref{algo: sampling pairs 2} finds a pair $(C_{j'},C_j)$ such that $\langle \Pi_{\ast,C_{j'}},\Pi_{\ast,C_j}\rangle \ge 2^{-\ell}-\kappa\eps$ with probability $\Omega(\min\{\epsilon^{4{\delta'}+2\gamma} d^2/m,1\}) \ge \delta''$ for some absolute constant $\delta'' > 0$, provided that $K_1\delta\geq 2\gamma$.
    
    Recall that $2^{-\ell-\ell'}\ge 8\eps$ and $2^{-\ell'}\le 1$, we conclude that there is a pair of columns of $\Pi V$ whose inner product is at least $(8-\kappa)\eps / 2^{-\ell'}$ with probability at least $\delta''$. This contradicts \Cref{lemma: inner product implies anticoncentration} since $\Pi$ is an $(\eps,2\delta/\gamma)$-subspace-embedding for $U\sim \cD_{2^{-\ell'}}$, whenever $2\delta/\gamma < \delta''/6$, so we let $\gamma\triangleq 12.5\delta/\delta''$ and $K_1\triangleq 25/\delta''$.
\end{proof}

Setting $L\triangleq s$, we can obtain the following generalized lower bound.
\begin{theorem}\label{thm: general}
There exist absolute constants $\eps_0, \delta_0, K_0, K_1 > 0$ such that the following holds.
For all $\eps \in (0,\eps_0)$, $\delta \in (0,\delta_0)$, $d\ge 1/\eps^2$, $s\leq 1/(9\eps)$ and $n\ge K_0  d^{2}/(\epsilon^2\delta)$, if
\begin{enumerate*}[label=(\roman*)]
    \item the column sparsity of $\Pi$ is $s$, and
    \item $\Pi$ is an $(\eps,\delta)$-subspace-embedding for $U\sim \widetilde{\cD}$,
\end{enumerate*}
then $\Pi$ must have $\Omega((\log^{-4} s)s^{-K_1\delta} d^2)$ rows.
\end{theorem}

\bibliography{refs}

\newcommand{\etalchar}[1]{$^{#1}$}
\begin{thebibliography}{CEM{\etalchar{+}}15}

\bibitem[ABTZ14]{ABTZ}
Haim Avron, Christos Boutsidis, Sivan Toledo, and Anastasios Zouzias.
\newblock Efficient dimensionality reduction for canonical correlation
  analysis.
\newblock {\em {SIAM} J. Sci. Comput.}, 36(5), 2014.

\bibitem[BZMD15]{BZMD}
Christos Boutsidis, Anastasios Zouzias, Michael~W. Mahoney, and Petros Drineas.
\newblock Randomized dimensionality reduction for k-means clustering.
\newblock {\em {IEEE} Trans. Inf. Theory}, 61(2):1045--1062, 2015.

\bibitem[CEM{\etalchar{+}}15]{Cohen15}
Michael~B. Cohen, Sam Elder, Cameron Musco, Christopher Musco, and Madalina
  Persu.
\newblock Dimensionality reduction for k-means clustering and low rank
  approximation.
\newblock In Rocco~A. Servedio and Ronitt Rubinfeld, editors, {\em Proceedings
  of the Forty-Seventh Annual {ACM} on Symposium on Theory of Computing, {STOC}
  2015, Portland, OR, USA, June 14-17, 2015}, pages 163--172. {ACM}, 2015.

\bibitem[Coh16]{Cohen16}
Michael~B. Cohen.
\newblock Nearly tight oblivious subspace embeddings by trace inequalities.
\newblock In {\em Proceedings of the Twenty-Seventh Annual ACM-SIAM Symposium
  on Discrete Algorithms}, SODA '16, pages 278--287, USA, 2016. Society for
  Industrial and Applied Mathematics.

\bibitem[CW17]{CW17}
Kenneth~L. Clarkson and David~P. Woodruff.
\newblock Low-rank approximation and regression in input sparsity time.
\newblock {\em Journal of the {ACM}}, 63(6):1--45, feb 2017.

\bibitem[NN13a]{NN13:OSNAP}
J.~{Nelson} and H.~L. {Nguyen}.
\newblock Osnap: Faster numerical linear algebra algorithms via sparser
  subspace embeddings.
\newblock In {\em 2013 IEEE 54th Annual Symposium on Foundations of Computer
  Science}, pages 117--126, 2013.

\bibitem[NN13b]{NN13:LB}
Jelani Nelson and Huy~L. Nguyen.
\newblock Sparsity lower bounds for dimensionality reducing maps.
\newblock In Dan Boneh, Tim Roughgarden, and Joan Feigenbaum, editors, {\em
  Symposium on Theory of Computing Conference, STOC'13, Palo Alto, CA, USA,
  June 1-4, 2013}, pages 101--110. {ACM}, 2013.

\bibitem[NN14]{NN14}
Jelani Nelson and Huy~L. Nguyen.
\newblock Lower bounds for oblivious subspace embeddings.
\newblock In Javier Esparza, Pierre Fraigniaud, Thore Husfeldt, and Elias
  Koutsoupias, editors, {\em Automata, Languages, and Programming}, pages
  883--894, Berlin, Heidelberg, 2014. Springer Berlin Heidelberg.

\bibitem[Woo14]{W14}
David~P. Woodruff.
\newblock Sketching as a tool for numerical linear algebra.
\newblock {\em Found. Trends Theor. Comput. Sci.}, 10(1-2):1--157, 2014.

\end{thebibliography}
\bibliographystyle{alpha}

\end{document}